\documentclass[%
 reprint,,bibtex,
superscriptaddress,
nofootinbib,
 amsmath,amssymb,
 aps,
prb,
]{revtex4-1}
\usepackage{units}
\usepackage{graphicx}
\usepackage{dcolumn}
\usepackage{bm}

\newcommand{\ved}{\varepsilon_{2D}}
\newcommand{\lk}{\left(}
\newcommand{\rk}{\right)}

\newcommand{\lK}{\left[}
\newcommand{\rK}{\right]}
\newcommand{\bE}{\mathbf E}
\newcommand{\bH}{\mathbf H}
\newcommand{\bk}{\mathbf k}
\newcommand{\bq}{\mathbf q}
\begin{document}

\preprint{APS/123-QED}

\title{Tunable THz Surface Plasmon Polariton based on Topological Insulator-Layered
Superconductor Hybrid Structure
}

\author{Mingda Li}
\email{Author to whom correspondence should be addressed: mingda@mit.edu}
\affiliation{Department of Nuclear Science and Engineering, Massachusetts Institute of Technology, 77 Massachusetts Avenue,
Cambridge, MA 02139, USA}

\author{Zuyang Dai}%
\affiliation{%
Department of Physics, Tsinghua University, Beijing 10084, China
}%

\author{Wenping Cui}
\affiliation{Institut f\"ur Theoretische Physik, Universit\"at zu K\"oln, Z\"ulpicher Str. 77, D-50937 K\"oln, Germany}

\author{Zhe Wang}
\affiliation{Department of Nuclear Science and Engineering, Massachusetts Institute of Technology, 77 Massachusetts Avenue, Cambridge, MA 02139, USA}

\author{\\Ferhat Katmis}
\affiliation{Department of Physics, Massachusetts Institute of Technology, 77 Massachusetts Avenue, Cambridge, MA 02139, USA}

\author{Peisi Le}
\affiliation{Department of Nuclear Science and Engineering, Massachusetts Institute of Technology, 77 Massachusetts Avenue,
Cambridge, MA 02139, USA}

\author{Jiayue Wang}
\affiliation{
Department of Engineering Physics, Tsinghua University, Beijing 10084, China
}

\author{Lijun Wu}
\affiliation{
Department of Condensed Matter Physics, Brookhaven National Lab, Upton, New York 11973, USA
}%
\author{Yimei Zhu}
\affiliation{
Department of Condensed Matter Physics, Brookhaven National Lab, Upton, New York 11973, USA
}%
\date{\today}

\begin{abstract}
We theoretically investigate the surface plasmon polariton (SPP) at the interface between 3D strong
topological insulator (TI) and layered superconductor-magnetic insulator structure. The tunability of SPP
through electronic doping can be enhanced when the magnetic permeability of the layered structure
becomes higher. When the interface is gapped by superconductivity or perpendicular magnetism, SPP
dispersion is further distorted, accompanied by a shift of group velocity and penetration depth. Such a shift
of SPP reaches maximum when the magnitude of Fermi level approaches the gap value, and may lead to
observable effects. The tunable SPP at the interface between layered superconductor and magnetism
materials in proximity to TI surface may provide new insight in the detection of Majorana Fermions.
\end{abstract}

\pacs{73.20.Mf, 73.25.+i}
\keywords{Surface Plasmon, Topological Insulator, Majorana Fermion}
\maketitle


\section{Introduction}
Surface plasmon polariton (SPP) is the collective excitation
of electrons at the interface between conductor and
dielectrics driven by electromagnetic (EM) waves. \cite{1pitarke2007theory, 2maier2007plasmonics}
Despite its wide applications in nanophotonics \cite{3nanophotonics2007springer}, near-
field optics and tip-enhanced Raman spectroscopy \cite{4kawata2001near, 5cui2014near},
and biological sensors and antennas \cite{6homola2006surface, 7becker2012plasmons}, SPP in general
suffers from problem of huge non-radiative loss due to the
strong absorption of the metal \cite{1pitarke2007theory, 2maier2007plasmonics, 3nanophotonics2007springer,
4kawata2001near} accompanied with
additional radiative loss \cite{8vary2010propagation}, which limits SPP's lifetime
and propagation length for further application in integrated
devices.

In order to solve the SPP loss problem, the low-loss
plasmonics based on graphene \cite{9wunsch2006dynamical, 10hwang2007dielectric, 11jablan2009plasmonics, 12ju2011graphene, 13koppens2011graphene, 14yan2012plasmonics, 15grigorenko2012graphene,
16maier2012graphene, 17bao2012graphene, 18tassin2012comparison,
19brar2013highly, 20jin2013terahertz, 21lu2013flexible, 22buljan2013graphene, 23Avouris} and topological
insulators (TI) \cite{24efimkin2012collective, 25efimkin2012spin, 26roslyak2013plasmons,
27schutky2013surface,28di2013observation, 29okada2013topological, 30stauber2013spin, 31chen2013tunable} has attracted much recent attention.
In far infrared and THz range, the major loss mechanism in
graphene lies in the scattering between electrons and optical
phonons \cite{11jablan2009plasmonics}. A number of studies in graphene plasmonics
have been conducted utilizing the properties of low loss and
tunability. \citeauthor{32yan2012tunable} \cite{32yan2012tunable} have reported enhanced plasmon
resonance in patterned graphene-insulator stack structure
comparing with single-layered graphene, while Ju et al \cite{12ju2011graphene}
demonstrated an enhanced tunability at THz range in
micro-ribbon graphene metamaterials. These efforts target
at manipulation of photons and miniaturization of optical
devices, and could be further integrated and hybridized
toward further applications in detectors, modulators or
other integrated devices.

On the other hand, in doped 3D TI, the electron-impurity
scattering becomes dominant due to weak electron-phonon
coupling \cite{28di2013observation}, with a further reduced backscattering
probability thanks to topological protected surface states
\cite{33bernevig2013topological, 34hasan2010colloquium, 35qi2011topological}. However, unlike the booming studies in graphene
plasmonics, the plasmon hybrid devices in TI have been
seldom reported, even with comparable performance in
THz range as well as additional features such as net spin
polarization, i.e. ``spin plasmon'' \cite{24efimkin2012collective, 25efimkin2012spin, 26roslyak2013plasmons, 31chen2013tunable} and spin-charge separation \cite{30stauber2013spin}.

Therefore, in this paper, we propose a plasmonic hybrid
structure composed of 3D TI in close contact with layered
superconductor. This structure provides a new platform where SPP waves are supported.
The tunability of the SPP propagation can be achieved
independently through either gate voltage and magnetic field. Since the Majorana bound states, which are an non-Abelian anyons in superconductors and may have great impact in topological quantum computation\cite{34hasan2010colloquium}, may exist at the boundary of 3D TI and superconductor, this plasmonic structure may provide a new perspective in search for Majorana bound states.

\section{Theory}
\subsection{Dispersion Relation of Anisotropic SPP Wave}
\begin{figure}
\begin{center}
\includegraphics[width=8cm]{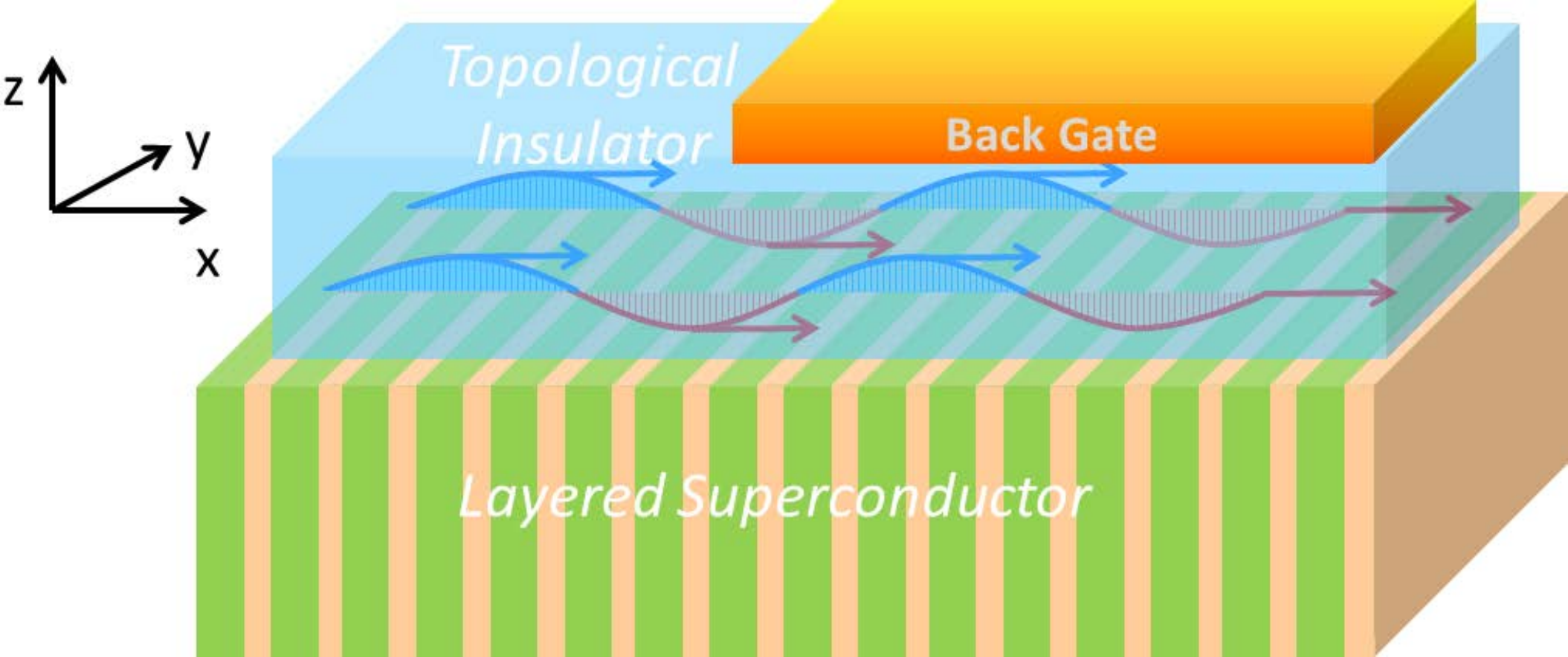}
\end{center}
\caption{The schematic configuration of 3D TI-layered
superconductor hybrid structure. The $z<0$ region consist of
alternating layers of superconductor and nonmagnetic or
magnetic insulator. The SPP wave propagates along the interface.
A back gate is present to tune the Fermi level of the interfacial
electrons, which leads to a change of dielectric functions and
furthermore a change of SPP propagation properties.}
\label{fig1}
\end{figure}

Since the SPP wave is localized along the interface, and
the Dirac electrons only exist on the surface of TI, we use
anisotropic dielectric functions to model the dielectric
function of a 3D TI in order to capture both the Dirac
dielectric functions of 2D chiral Dirac electrons at the
surface (Fig. \ref{fig1}, $xy$ plane, $z=0$) as well as the dielectric
constant in the bulk (Fig. \ref{fig1}, $z>0$ region).

In order to describe wave propagation in the layered
structure( Fig. \ref{fig1} $z<0$ region), we adopt the method proposed
by Averkov et al \cite{36PhysRevB.87.054505} using anisotropic dielectric function.
This is valid when $\lambda_{SP} \gg D$ , where $\lambda_{SP}$ is the SPP
wavelength, and $D$ is the spatial periodicity of the layered
structure. Since we are interested in long wavelength THz
range, the condition that $\lambda_{SP} \gg D$ is guaranteed to meet.
In this approach, the anisotropy leads to the existence of
optical axis where EM wave suffers no birefringence \cite{37hecht1998optics}.
Thus the electric field and magnetic field at the interface
can be written as a superposition of ordinary wave and
extraordinary wave:
\begin{eqnarray}\label{eq1}
\begin{cases}
\mathbf{E}_j=\lk \bE_{j}^{o}e^{-\kappa_j^{o}|z|}+\bE_{j}^{e}e^{-\kappa_j^{e}|z|}\rk e^{i(q_xx+q_yy-\omega t)}\\\\
\mathbf{H}_j=\lk \bH_{j}^{o}e^{-\kappa_j^{o}|z|}+\bH_{j}^{e}e^{-\kappa_j^{e}|z|}\rk e^{i(q_xx+q_yy-\omega t)}
\end{cases}
\end{eqnarray}
where $j=1, 2$ is for the TI and layered superconductor
side, respectively. Based on our model, the TI dielectric
function is defined as $\varepsilon_1=\lk\varepsilon_{2D}(q,\omega), \varepsilon_{2D}(q,\omega), \varepsilon_{d}\rk$ with
optical axis along $z$ direction, while the dielectric function of
layered superconductor is defined as
$\varepsilon_2=\lk\varepsilon_{c}(q,\omega), \varepsilon_{ab}(q,\omega), \varepsilon_{ab}(q,\omega)\rk$,
which is anisotropic along $x$ direction.

By noticing the different direction of optical axis in the
upper TI region and lower superconductor region, that for
ordinary wave and extraordinary wave we have
$E_{1z}^{o}=0$, $E_{2x}^{o}=0$, and $H_{1z}^{e}=0$, $H_{2x}^{e}=0$, respectively. The $x$
components of the EM fields are shown in Appendix \ref{appendix1}, with
light speed in vacuum $c=1$ and $\mu$ is the magnetic
permeability of the layered superconductor material.

Substituting the EM field components back to eq. (\ref{eq1}), we
obtain the localization constants, i.e. inverse of penetration
depth away from the interface.
\begin{eqnarray}
\kappa_1^{o}&=&\sqrt{q_x^2+q_y^2-\varepsilon_{2D}(q,\omega)\omega^2}\\
\kappa_1^{e}&=&\sqrt{\varepsilon_{2D}(q,\omega)\lk \tfrac{q_x^2+q_y^2}{\varepsilon_{d}}-\omega^2 \rk}\\
\kappa_2^{o}&=&\sqrt{q_x^2+q_y^2-\mu\varepsilon_{ab}(q,\omega)\omega^2}\\
\kappa_2^{e}&=&\sqrt{\tfrac{\varepsilon_{c}(q,\omega)}{\varepsilon_{ab}(q,\omega)}q_x^2+q_y^2-\mu\varepsilon_{c}(q,\omega)\omega^2}
\end{eqnarray}
The dispersion relation of the resulting surface wave can be
written in a determinant form:
\begin{equation}\label{matirx3}
\begin{vmatrix}
-i\kappa_1^{o} & \frac{i\omega^2\varepsilon_{1}}{\kappa_1^{e}}&-\frac{q_y^2-\lk\kappa_2^{o} \rk^2}{i\mu\kappa_2^{o}} &  0\\\\
-q_y^2 & q_x^2        &      0    &  -q_x^2+\mu\omega^2\varepsilon_{ab}\\\\
\kappa_1^{o}q_y^2 & \frac{\omega^2q_x^2\varepsilon_{1}}{\kappa_1^{e}}&\frac{q_x^2q_y^2}{\mu\kappa_2^{o}} &  \omega^2\varepsilon_{ab}\kappa_2^{e}\\\\
1&1&-1&-1
\end{vmatrix}
=0
\end{equation}
In this paper, we only consider the wave propagating
along x direction, neglecting the oblique excitation. The
dispersion relation can finally be simplified as
\begin{equation}\label{dispersion}
 q=\omega\sqrt{\frac{\varepsilon_{d}\varepsilon_{ab}(q,\omega)\lK \varepsilon_{c}(q,\omega)-\mu\varepsilon_{2D}(q,\omega)\rK}
 {\varepsilon_{c}(q,\omega)\varepsilon_{ab}(q,\omega)-\varepsilon_{2D}(q,\omega)\varepsilon_{d}}}
\end{equation}
which is the main analytical result. In this expression, $\mu$ is
the effective permeability of the layered superconductor
structure, q is the wavenumber along $x$ direction, and can be
a complex number. When both the upper and lower
materials are isotropic, i.e. $\varepsilon_{2D}=\varepsilon_{d}=\varepsilon_{1}$
and $\varepsilon_{ab}=\varepsilon_{c}=\varepsilon_{2}$,
it is further reduced to the well-known result
$q=\omega\sqrt{\tfrac{\varepsilon_{1}\varepsilon_{2}}{\varepsilon_{1}+\varepsilon_{2}}}$.

\subsection{Dielectric Function of Layered Superconductor}
Define dimensionless frequency $\Omega=\omega/\omega_J$, with $\omega_J$ the
Josephson plasmon frequency of the layered
superconductor, the dielectric function of layered
superconductor can be written as \cite{36PhysRevB.87.054505}:
\begin{equation}\label{eq5Omega}
\varepsilon_{c}(\Omega)=\varepsilon_{s}\lk1-\frac{1}{\Omega^2}\rk, \quad
\varepsilon_{ab}(\Omega)=\varepsilon_{s}\lk1-\frac{\gamma^2}{\Omega^2}\rk
\end{equation}
where the imaginary parts are neglected. Throughout this
article, we take the value reported in \cite{36PhysRevB.87.054505} and set the
interlayer dielectric constant $\varepsilon_{s}=16$, current-anisotropy
parameter $\gamma=200$ and $\omega_J=\unit[4]{meV}$.

\subsection{Dielectric Function of Gapless Topological Insulator}
The dielectric function of the chiral gapless topological
insulator surface have been reported in \cite{24efimkin2012collective, 25efimkin2012spin, 27schutky2013surface}, where
for the Hamiltonian for 2D helical Dirac electron gas $H_0$:
\begin{equation}
 H_0=\hbar v_F\sum_{k}\Psi_k^+\lk\hat{z}\times\vec{k}\rk\cdot \vec{\sigma}\Psi_k
\end{equation}
the Lindhard Dielectric function $\ved(q,\omega)$ can be written as
\begin{equation}\label{epsion7}
 \ved(q,\omega)=1-V_{2D}(q)\Pi(q,\omega)=1-\frac{e^2}{2\varepsilon_{0}q}\Pi(q,\omega)
\end{equation}
with the polarization operator
\begin{widetext}
\begin{equation}\label{Pieq8}
\Pi(q,\omega)=\frac{g}{4\pi^2}\sum_{\gamma, \gamma'}\int d\bk\frac{n_{\bk, \gamma}-n_{\bk+\bq, \gamma'}}
{\hbar\omega+E_{\bk, \gamma}-E_{\bk+\bq, \gamma'}+i\delta}
|\left\langle f_{\bk, \gamma}|f_{\bk+\bq, \gamma'} \right\rangle|^2
\end{equation}
\end{widetext}
In this expression, $g$ is the spin/valley degeneracy, for the
chiral states we have $g=1$ due to spin-momentum locking,
$n_{\bk, \gamma}$ is Fermi occupation value at energy eigenvalue,
$E_{\bk, \gamma}=\gamma\hbar v_Fk$, with $\gamma=1$ for conduction band and $\gamma=-1$
for valence band, respectively. The spinor eigenstates
$|f_{\bk, \gamma} \rangle=\lk e^{-i\theta_\bk /2},~ i\gamma e^{i\theta_\bk /2}  \rk/\sqrt{2}$,
with $\theta_\bk $ defined as
$\tan \theta_\bk=k_y/k_x$. We take the value of Fermi velocity
$v_F=\unit[6.2\times10^5]{m/s}$ \cite{38zhang2009topological} in all calculations.

Eq. (\ref{Pieq8}) can be calculated either analytically \cite{27schutky2013surface, 39hwang2007dielectric}or
numerically, thanks to the identical expression of
polarization operator within RPA for simple Dirac gas and
helical Dirac gas. In our approach, we choose $T=\unit[10]{K}$ for
numerical integration to avoid discontinuity, and compare
with analytical result with at least 6 digit agreement. Then
the numerical result for gapless TI is applied to solve for
the dielectric function of TI when surface state is gapped.

\subsection{Dielectric Function of Gapped Topological Insulator}
Either the magnetic field perpendicular to the interface
(along $z$ direction in Fig. \ref{fig1}) or the superconductivity would
open up a gap to the gapless Dirac cone, and lift out the
degeneracy at $k=0$, and further alter the dielectric function
as well SPP dispersions. In order to take the effect of gaps
into account, we adopt the Bogoliubov-de Gennes
Hamiltonian \cite{33bernevig2013topological, 40fu2008superconducting}, which can be regarded as a
generalization of eq. (\ref{eq5Omega}):
\begin{widetext}
\begin{equation}
H_{BdG}\!\!=\!\!\frac{1}{2}\sum_\bk \Psi_\bk^+
\left( \begin{array}{cc}
\!\!k_x\sigma_y\!\!-\!\!k_y\sigma_x\!\!+\!\!M\sigma_z\!\!-\!\!E_F\!\!\!\! &\!\!\!\!i|\Delta|\sigma_y\!\!\\
\!\!-i|\Delta|\sigma_y \!\!\!\!&\!\!\!\!-k_x\sigma_y\!\!-\!\!k_y\sigma_x\!\!-\!\!M\sigma_z\!\!-\!\!E_F\!\!
\end{array} \right)
\Psi_\bk
\end{equation}
\end{widetext}
where spinor $\Psi_\bk =\lk c_{\bk\uparrow}, c_{\bk\downarrow},
c_{-\bk\uparrow}^+, c_{-\bk\downarrow}^+  \rk$ and denote
the magnetic gap and superconductivity gap, respectively,
$E_F$ is the chemical potential. Both eq. (\ref{epsion7}) and eq. (\ref{Pieq8}) still
hold, but the eigenvalues and eigenvectors are changed.
The eigenvalues can now be written as
\begin{eqnarray*}
E_1\!\!&=&\!\!\sqrt{|\Delta|^2\!\!+\!\!2\lk|\Delta|^2M^2\!\!+\!\!E_F^2M^2+\!\!E_F^2k^2 \rk^{1/2}\!\!+\!\!M^2\!\!+\!\!k^2\!\!+\!\!E_F^2}\\
E_2\!\!&=&\!\!\sqrt{|\Delta|^2\!\!-\!\!2\lk|\Delta|^2M^2\!\!+\!\!E_F^2M^2+\!\!E_F^2k^2 \rk^{1/2}\!\!+\!\!M^2\!\!+\!\!k^2\!\!+\!\!E_F^2}\\
E_3\!\!&=&\!\!-\sqrt{|\Delta|^2\!\!-\!\!2\lk|\Delta|^2M^2\!\!+\!\!E_F^2M^2+\!\!E_F^2k^2 \rk^{1/2}\!\!+\!\!M^2\!\!+\!\!k^2\!\!+\!\!E_F^2}\\
E_4\!\!&=&\!\!-\sqrt{|\Delta|^2\!\!+\!\!2\lk|\Delta|^2M^2\!\!+\!\!E_F^2M^2+\!\!E_F^2k^2 \rk^{1/2}\!\!+\!\!M^2\!\!+\!\!k^2\!\!+\!\!E_F^2}\\
\end{eqnarray*}
The corresponding unnormalized eigenvectors are shown in
Appendix \ref{appendix1}.

When the surface state is gapped, the EM constitutive
relations have been modified with an additional axion term
\cite{41karch2011surface} and leads to electromagnetic effect. However, it is
shown that such effect \cite{41karch2011surface, 42li2013topological} only has $\sim\alpha^2$ correction to
the SPP dispersion, where $\alpha$ is the fine structure constant,
thus in the present calculation we neglect the effect from
axion electrodynamics

\section{Results and Discussions}
In order to conveniently express the dielectric functions
for both TI and layered superconductor, we define
dimensionless wavenumber $Q=v_Fq_x/(c\omega_J)$, and all
energies are dimensionless and expressed in the unit of $\omega_J$.
\begin{figure}
\begin{center}
\includegraphics[width=8.78cm]{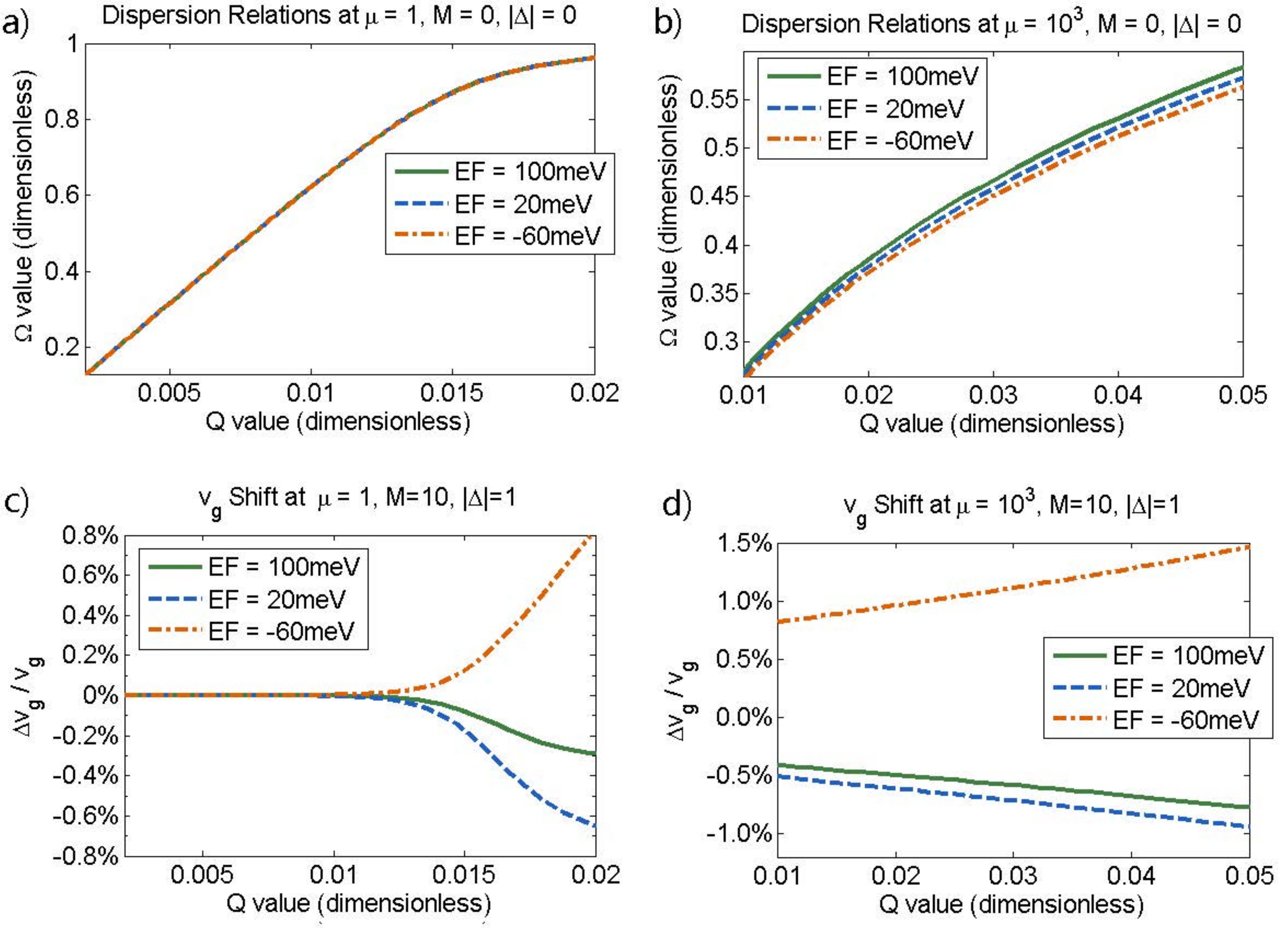}
 \end{center}
\caption{The dispersion relations (a-b) and gap-induced group
velocity changes (c-d) of SPP at various Fermi levels with respect
to gap opening. The Fermi levels are taken at 3 different values
with $\unit[80]{meV}$ interval. In c), at $\mu=1$, even if the tuning of Fermi
level does not change much to the dispersion relation itself as
shown in a), it still shows a shift to SPP group velocity at high $Q$
range.}
   \label{fig2}
\end{figure}

\begin{figure}
\begin{center}
\includegraphics[width=8.78cm]{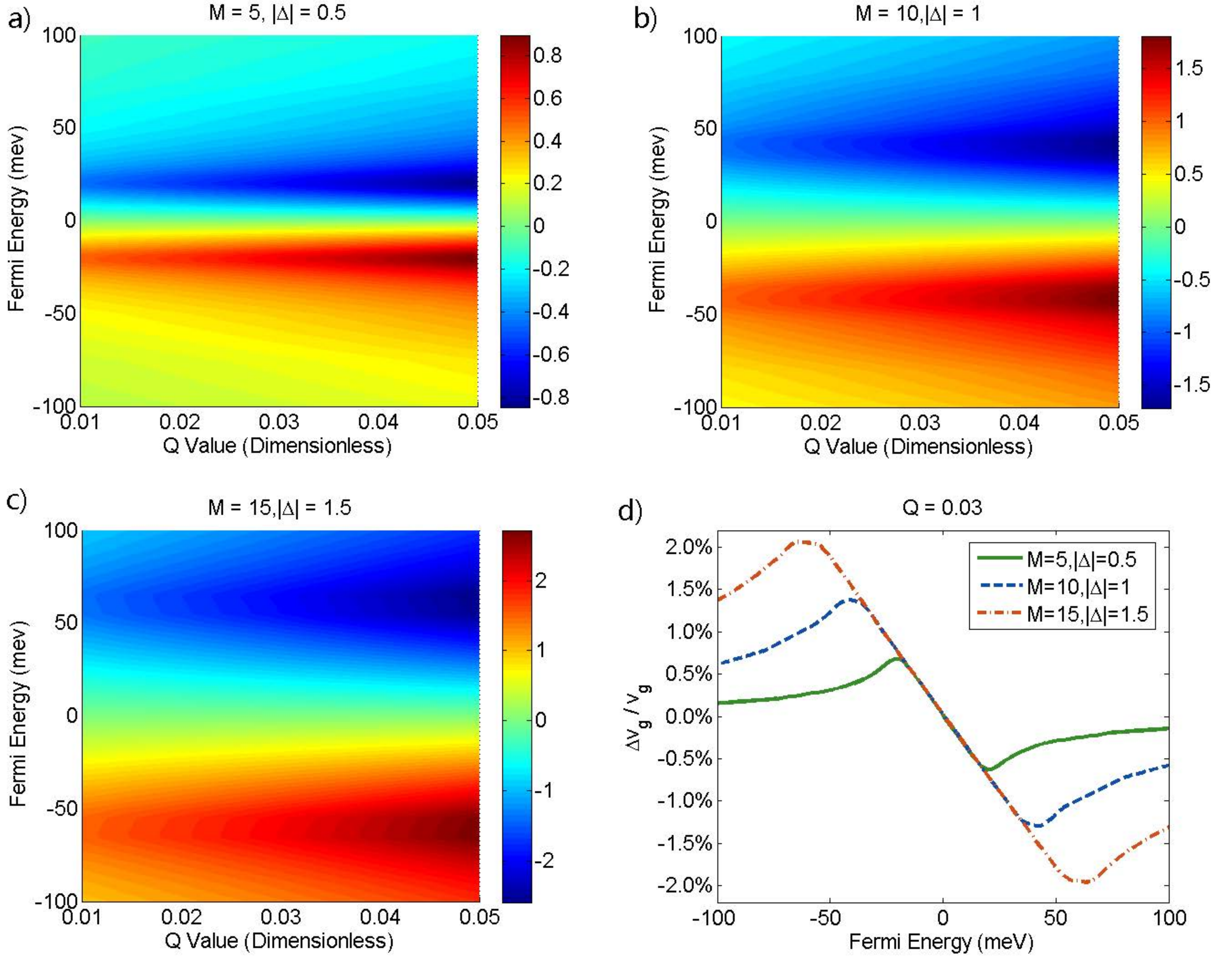}
 \end{center}
\caption{The percentage shift of SPP group velocity as a function of
$Q$ and $E_F$, at 3 different gap values $M=5$ $|\Delta|=0.5$ (a),
$M=10$ $|\Delta|=1.0$, (b) and $M=15$ $|\Delta|=1.5$, (c).
The change is negative for
$E_F>0$ and positive for $E_F<0$, and reaches maximum when the
Fermi level is close to the gap values (d). Notice all gap values are
expressed in the unit of $\omega_J$ ($\omega_J=\unit[4]{meV}$ throughout calculation),
for instance the green solid line $M=5$ $|\Delta|=0.5$ corresponds to
$M=\unit[20]{meV}$ $|\Delta|=\unit[2]{meV}$.}
  \label{fig3}
\end{figure}
The SPP dispersion relations are obtained by solving Eq. (\ref{dispersion}).
The typical dispersion relations at various gate
voltages, i.e. Fermi levels $E_F$, are shown in Fig. \ref{fig2} a and b.
We see an enhanced tunability, i.e. shift of dispersion
relation when $E_F$ varies, when the lower layer has increased
effective magnetic permeability $\mu$. By defining the SPP
group velocity $v_g=d\Omega/dQ$, we see an additional change
of propagation properties induced by either $M$ or $|\Delta|$ (Fig. \ref{fig2}
c and d) by plotting the percentage SPP group velocity shift
$$\frac{\Delta v_g}{v_g}=\frac{v_g(gapped)-v_g(gapless)}{v_g(gapless)}$$
Here both the magnetic
gap $M$ and superconductivity gap $|\Delta|$ are dimensionless in
the unit of Josephson plasmon frequency $\omega_J$.

In order to further investigate the effect of gap to the SPP
propagation, the relative shift of group velocity $\Delta v_g/v_g$  as a
function of dimensionless wavenumber $Q$ and Fermi level
$E_F$ are shown in Fig. \ref{fig3}, at different values of gap. It can be
seen directly that the shift is increased at larger gap value,
and it increases as a function of Q. Most importantly, the
shift reaches peak value when the Fermi level approaches
to the gap value.
This feature can be seen more clearly in a line plot with
fixed $Q$ value (Fig. \ref{fig3} d).We also see that SPP group velocity
$v_g$  does not shift when Fermi level $E_F=0$.
\begin{figure}
\begin{center}
\includegraphics[width=8.78cm]{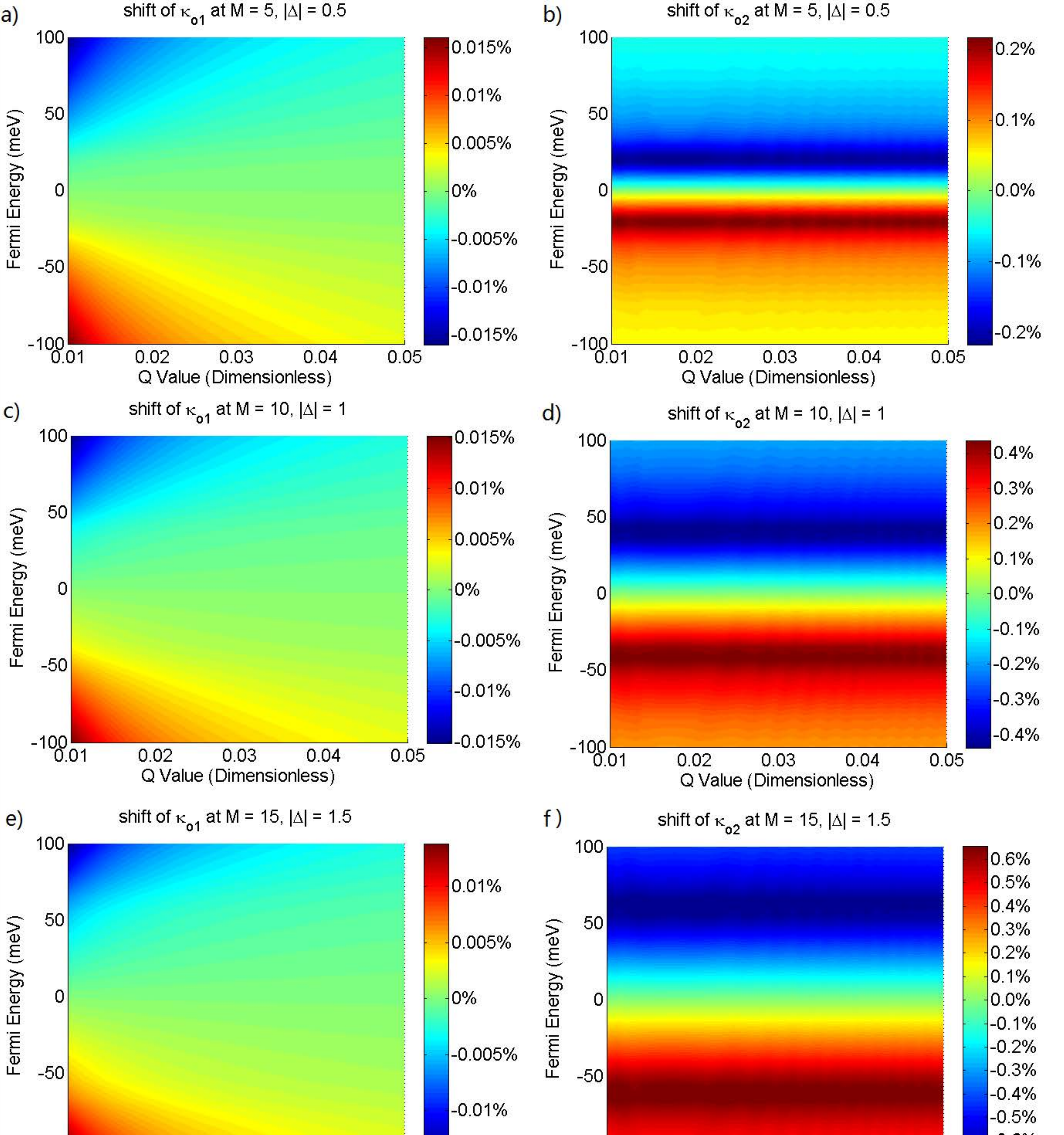}
 \end{center}
\caption{Relative shift of o-light component of localization constants
for TI $\kappa_1^{(o)}$ (figs a, c, e), and layered superconductor $\kappa_2^{(o)}$
(figs b, d, f). The localization constants for e-light $\kappa_1^{(e)}$
and $\kappa_2^{(e)}$ have
different magnitudes but similar feature of  $\kappa_1^{(o)}$ and $\kappa_2^{(o)}$,
respectively. We see that the shift is not sensitive to Fermi level
$E_F$ of topological insulators, but always reaches peak values when
$E_F$ is near the value of gap. Unlike the shift of group velocity,
which is $Q$ dependent, the localization constants are almost
independent $Q$ value.}
 \label{fig4}
\end{figure}
In contrary to the $Q$ and $E_F$ dependent shift of group
velocity caused by energy gap, the localization constants
show a qualitatively different behavior. The localization
constant in TI side $\kappa_1^{(o)}$ is neither sensitive to $E_F$ nor to $Q$
(Fig. \ref{fig4} a, c, e), while in layered superconductor side, $\kappa_2^{(o)}$ is
tunable by $E_F$ but still $Q$-independent. The importance of
the localization constant shift as $Q$ can hardly be
overestimated, in that it indicates that the different
propagation properties of SPP is almost fully coming from
the group velocity, other than the difference in SPP
wavelength. The fewer controllable variables would
indefinitely make the experimental results easier to explain.

The reason that the maximum $v_g$ shift occurs when the
gap matches the Fermi level can be understood in the light of
electronic transition and available electron density for
plasmonic oscillation. Electrons in the conduction band are
mostly extended and dominate the surface plasmon
excitation. When the Fermi level is lower than the energy
gap, charge carriers need additional energy to fulfill the
transition to the conduction band in order to contribute to
the collective excitation. On the other hand, since the SPP
is excited by EM waves, when the Fermi level is higher
than the energy gap, the electronic transition by optical
excitation is limited by occupied electrons. The increasing
forbidden transition leads to a reduced shift of dielectric function, and
moreover the $v_g$ shift is reduced accordingly. Thus, only
when the Fermi level reaches the value of energy gap are
the number of available electrons participating in the
electronic transition and collective excitation maximized,
and leads to a shift of SPP frequency and group velocity.

\section{CONCLUSIONS}
We provide a generic theoretical framework to study the
surface plasmon polariton (SPP) at the interface between
topological insulator (TI) and layered superconductors. The
SPP in this hybrid structure may be widely applied to study
novel optical and transport phenomena at the interface,
through the tunability of SPP propagation by gating or
gapping the surface states of the TI. It can also be
generalized to a larger category of materials in proximity to
TI surface. For instance, \citeauthor{43PhysRevLett.110.186807} \cite{43PhysRevLett.110.186807,44orona2012fabrication} have
shown that when the ferromagnetic insulator EuS is on the
top of topological insulator Bi$_2$Se$_3$, it induces significant
magnetic moment in Bi$_2$Se$_3$ thin films and induces breaking
of T-reversal symmetry. The SPP in such Ferromagnetic-
TI hybrid structure can also be studied within this approach.

Furthermore, Majorana zero mode is predicted to exist
\cite{35qi2011topological, 40fu2008superconducting, 45fu2009probing} as a domain wall state at the interface between
TI and ferromagnetic-superconductor boundaries.
Therefore, at the TI-layered superconductor interface,
propagating SPP may interact with the Majorana domain
wall state, and leads to the shift of the SPP propagation
properties, including group velocity, reflectivity and
transmissivity, etc. In this regards, the change of optical
properties of SPP can be considered a semi-classical
manifestation of the existence of Majorana fermions, which
is a pure quantum phenomenon with non-Abelian statistics.
The existence of the zero mode level may contribute to
electronic transition, and leads to a further change of
dielectric functions. Unlike transport measurements, which
involve only a single Majorana fermion domain wall state,
this hybrid structure can be regarded as the interaction
between SPP and a series of domain wall states, coming
from each domain wall state along the interface. In a
nutshell, the SPP on the TI surface may provide insights to
the detection of Majorana Fermions as a conceptually novel
platform.
\acknowledgements
Author Mingda Li would thank Prof. Ju Li for his
generous support and helpful discussions.
\appendix
\section{Components of EM Fields}\label{appendix1}
$$H_{1x}^{(o)}=\frac{\kappa_1^{(o)}}{i\omega}E_{1y}^{(o)},
~E_{1x}^{(o)}=-\frac{q_y}{q_x}E_{1y}^{(o)},
~H_{1y}^{(o)}=-\frac{i\kappa_1^{(o)}q_y}{\omega q_x}E_{1y}^{(o)}$$

$$H_{1z}^{(o)}\!\!=\!\!\frac{q_x^2+q_y^2}{\omega q_x}E_{1y}^{(o)},
~H_{1x}^{(e)}\!\!=\!\!\frac{i\omega\varepsilon_{2D}(q, \omega)}{\kappa_1^{(e)}}E_{1y}^{(e)},
~E_{1x}^{(e)}\!\!=\!\!\frac{q_x}{q_y}E_{1y}^{(e)}$$
$$
H_{1y}^{(e)}\!\!=\!\!-\frac{i\omega q_x\varepsilon_{2D}(q, \omega)}
{q_y\kappa_1^{(e)}}E_{1y}^{(e)},
E_{1z}^{(e)}\!\!=\!\!\lk\!\!\frac{i\kappa_1^{(e)}}{q_y}+\frac{i\omega^2\varepsilon_{2D}(q, \omega)}
{q_y\kappa_1^{(e)}} \!\!\rk\!\! E_{1y}^{(e)}
$$
$$H_{2z}^{(o)}=\frac{q_x}{\omega\mu}E_{2y}^{(o)},
~H_{2y}^{(o)}=\frac{iq_xq_y}{\omega\mu\kappa_2^{(o)}}E_{2y}^{(o)},
~E_{2z}^{(o)}=-\frac{iq_y}{\kappa_2^{(o)}}E_{2y}^{(o)}$$
$$H_{2x}^{(o)}\!\!=\!\!\frac{q_y^2-\lk\kappa_2^{(o)}\rk^2}{i\omega\mu\kappa_2^{(o) }}E_{2y}^{(o)},
~E_{2z}^{(e)}\!\!=\!\!\frac{\kappa_2^{(e)}}{iq_y}E_{2y}^{(e)},
~H_{2z}^{(e)}\!\!=\!\!\frac{\omega\varepsilon_{ab}(q, \omega)}{q_x}E_{2y}^{(e)}$$
$$
H_{2y}^{(e)}\!\!=\!\!-\frac{\omega\varepsilon_{ab}(q, \omega)\kappa_2^{(e)}}
{iq_xq_y}E_{2y}^{(e)},~
E_{2x}^{(e)}\!\!=\!\!\frac{q_x^2-\mu\omega^2\varepsilon_{ab}(q, \omega)}
{q_xq_y} E_{2y}^{(e)}
$$
\section{(Un-normalized) Eigenvectors of BdG Hamiltonian}\label{appendix2}
Define $k^2=k_x^2+k_y^2$, $\delta^-=|\Delta|-M$ and $\delta^+=|\Delta|+M$,
the four eigenvectors can be written as:
\begin{eqnarray*}
|f_{\bk, 1} \rangle\!\!&=&\!\!\lk\!\!-1, \frac{\delta^-\!\!+\!\!\sqrt{(\delta^-)^2+k^2}}{ik_x+k_y}
, -\frac{\delta^-\!\!+\!\!\sqrt{(\delta^-)^2+k^2}}{ik_x+k_y}, 1\!\!\rk^T\\
|f_{\bk, 2} \rangle\!\!&=&\!\!\lk\!\!1, \frac{\delta^+\!\!-\!\!\sqrt{(\delta^+)^2+k^2}}{ik_x+k_y}
, \frac{\delta^+\!\!-\!\!\sqrt{(\delta^+)^2+k^2}}{ik_x+k_y}, 1\!\!\rk^T\\
|f_{\bk, 3} \rangle\!\!&=&\!\!\lk\!\!-1, \frac{\delta^-\!\!-\!\!\sqrt{(\delta^-)^2+k^2}}{ik_x+k_y}
, \frac{-\delta^-\!\!+\!\!\sqrt{(\delta^-)^2+k^2}}{ik_x+k_y}, 1\!\!\rk^T\\
|f_{\bk, 4} \rangle\!\!&=&\!\!\lk\!\!1, \frac{\delta^+\!\!+\!\!\sqrt{(\delta^+)^2+k^2}}{ik_x+k_y}
, \frac{\delta^+\!\!+\!\!\sqrt{(\delta^+)^2+k^2}}{ik_x+k_y}, 1\!\!\rk^T
\end{eqnarray*}
\bibliography{Bibliography}
\end{document}